\pdfoutput=1
\documentclass{article}
\usepackage{graphicx}
\usepackage[english]{babel}

\title{
\includegraphics[width=0.35\textwidth]{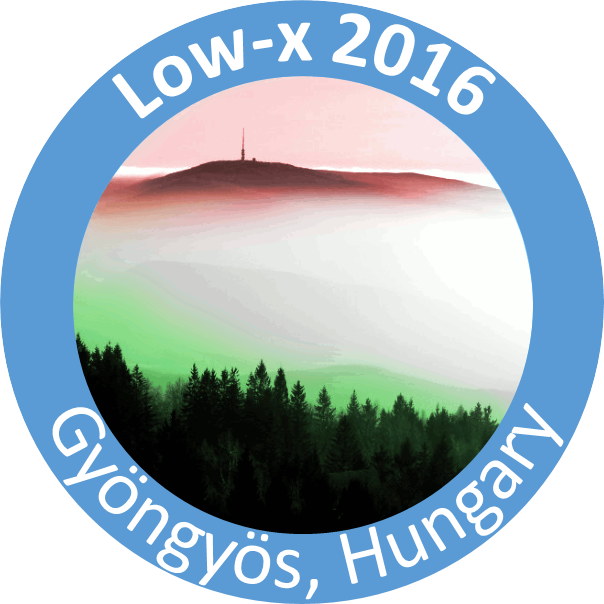}\\[1cm]
Measurement of the photon and jet production with the ATLAS detector}
\author{{Michal Svato\v s$^1$, On behalf of the ATLAS Collaboration}\\[1ex]
$^1$Institute of Physics, AV CR, Na Slovance 1999/2, Prague, Czech Republic\\
}

\begin{document}
\fontfamily{lmss}\selectfont
\maketitle

\begin{abstract}
Production of prompt isolated photons at hadron colliders provides a probe of perturbative QCD and can be used to study the gluon density function of the proton. The ATLAS collaboration has performed precise measurements of the inclusive production of isolated prompt photons in \mbox{20.2 fb${}^{-1}$} of data collected at \mbox{$\sqrt{s}=8$ TeV} and in \mbox{6.4 pb${}^{-1}$} of data collected at \mbox{$\sqrt{s}=13$ TeV}. The measurements are compared with state-of-the-art theory predictions at NLO in QCD and with predictions of several MC generators.\\
Production of inclusive jets at hadron colliders also provides a probe of perturbative QCD. The inclusive jet production cross-section was measured in \mbox{78 pb${}^{-1}$} of data collected at \mbox{$\sqrt{s}=13$ TeV}. Results have been compared with (state-of-the-art) theory predictions at NLO in QCD, interfaced with different parton distribution functions.
\end{abstract}

\section{Introduction}
Measurements of prompt photon production and of jet production provide a tool for testing perturbative Quantum Chromodynamics (pQCD). These measurements can be used also to study parton distribution functions (PDF). For example, Compton scattering of quarks and gluons ($qg\longrightarrow q\gamma$) can be used to study parton distribution function of the gluon. Improved understanding of these processes can lead to the improvement in other measurements. For example, photon production is a background for Higgs production in the diphoton channel. Jet production is a background for many SUSY/Exotics processes. Finally, precise jet cross-section measurements allow the extraction of information about the strong coupling constant $\alpha_{S}$ and probes QCD at scales sensitive to new physics.

\section{The ATLAS detector}
The ATLAS detector \cite{Aad:2008zzm} at LHC \cite{Evans:2008zzb} is a multi-purpose particle physics detector with a forward-backward symmetric cylindrical geometry and nearly 4$\pi{}$ coverage in solid angle. It consists of an inner tracking detector surrounded by a thin superconducting solenoid providing a 2 T axial magnetic field, electromagnetic and hadronic calorimeters, and a muon spectrometer. The inner tracking detector covers the pseudorapidity range $|\eta|<$ 2.5.

\section{Photon reconstruction, identification, isolation}\label{sec_ph_properties}
Photons are reconstructed from energy deposits in the electromagnetic calorimeter with transverse energy \mbox{$E_{T} >$ 2.5 GeV} in towers of $3\times5$ cells in $\eta \times \phi$.
\newline\indent Reconstructed photons are selected for analyses using identification and isolation criteria. Photon identification is used to reject hadronic background by applying requirements on the energy leaking into the hadronic calorimeter and on the shower-shape variables. The "tight" selection is optimised to reduce the contribution from jets with one or more hard $\pi^{0}$ decaying to photons and carrying most of the jet energy. Photon isolation is based on isolation energy ($E_{T}^{iso}$). A photon is considered isolated if its $E_{T}^{iso}$ is less than some isolation cut value. The $E_{T}^{iso}$ is based on the energy measured in topological clusters within a cone of radius R=0.4 with the core of the cone removed. In the photon measurements presented here, the cut is \mbox{4.8 GeV + 4.2$\times 10^{-3}\times E_{T}^{\gamma}$}.
\newline\indent Prompt photons, which represent the signal in the analysis, can belong to two different groups - direct photons and fragmentation photons. Direct photons originate from the hard processes. Fragmentation photons are emitted in the fragmentation of a high transverse momentum parton.

\section{Photon cross-section measurement at \mbox{$\sqrt{s}$=8 TeV}}\label{sec_ph_8tev}
The measurement of isolated prompt photon cross-section at \mbox{$\sqrt{s}$=8 TeV} \cite{Aad:2016xcr} is based on \mbox{20.2 fb${}^{-1}$} of data collected in 2012.
\newline\indent Events are selected according the following criteria: each event is required to have a reconstructed vertex consistent with the average beam-spot position, where the vertex is required to have at least two associated tracks. Photons are required to have $|\eta^{\gamma}|<1.37$, $1.56\leq |\eta^{\gamma}|<2.37$ and $25<E_{T}^{\gamma}<1500$ GeV. They are triggered using single photon triggers with thresholds 20, 40, 60, 80, 100, and 120 GeV. Data taken by these triggers are multiplied by appropriate prescales and combined together. Tight isolated photons are considered as signal here.
\newline\indent The background is subtracted using the so-called 2D sideband method. The purpose of this method is to remove the residual background from meson decays and jets. It is based on isolation and identification criteria. In this method, photons are divided into four regions (A: tight and isolated; B: tight and non-isolated, C: non-tight and isolated; D: non-tight and non-isolated). Regions B, C and D contain both signal and background events, region A contains only signal ones. Based on the numbers of photons in each category, the number of signal photons in data can be estimated.
\newline\indent Data are compared to Monte Carlo simulations. Pythia 8.165 \cite{Sjostrand:2007gs} (using CTEQ6L1 PDF), Sherpa 1.4.0 \cite{Gleisberg:2008ta} (using CT10 PDF), \textsc{JetPhox} \cite{Catani:2002ny} and \textsc{PeTer} \cite{Becher:2012xr} (both using CT10 PDF) are used. \textsc{JetPhox} is a parton-level generator for predictions of processes with photons in the final state. It achieves next-to-leading-order (NLO) accuracy for both direct and fragmentation photon processes. \textsc{PeTer} is a parton-level NLO generator which contains resummation of threshold logarithms in addition.
\newline\indent Systematic uncertainties are dominated by the energy scale uncertainty at the high-$E_{T}^{\gamma}$ region, by the uncertainty of a 2D sideband method and uncertainties of the content of direct and fragmentation photons in the low-$E_{T}^{\gamma}$ region. The luminosity uncertainty is 1.9\%. The statistical uncertainty is between 1 \% and 2 \% (except for high-$E_{T}^{\gamma}$ bins).
\begin{figure}[h]
\centering
	\includegraphics[width=0.75\textwidth]{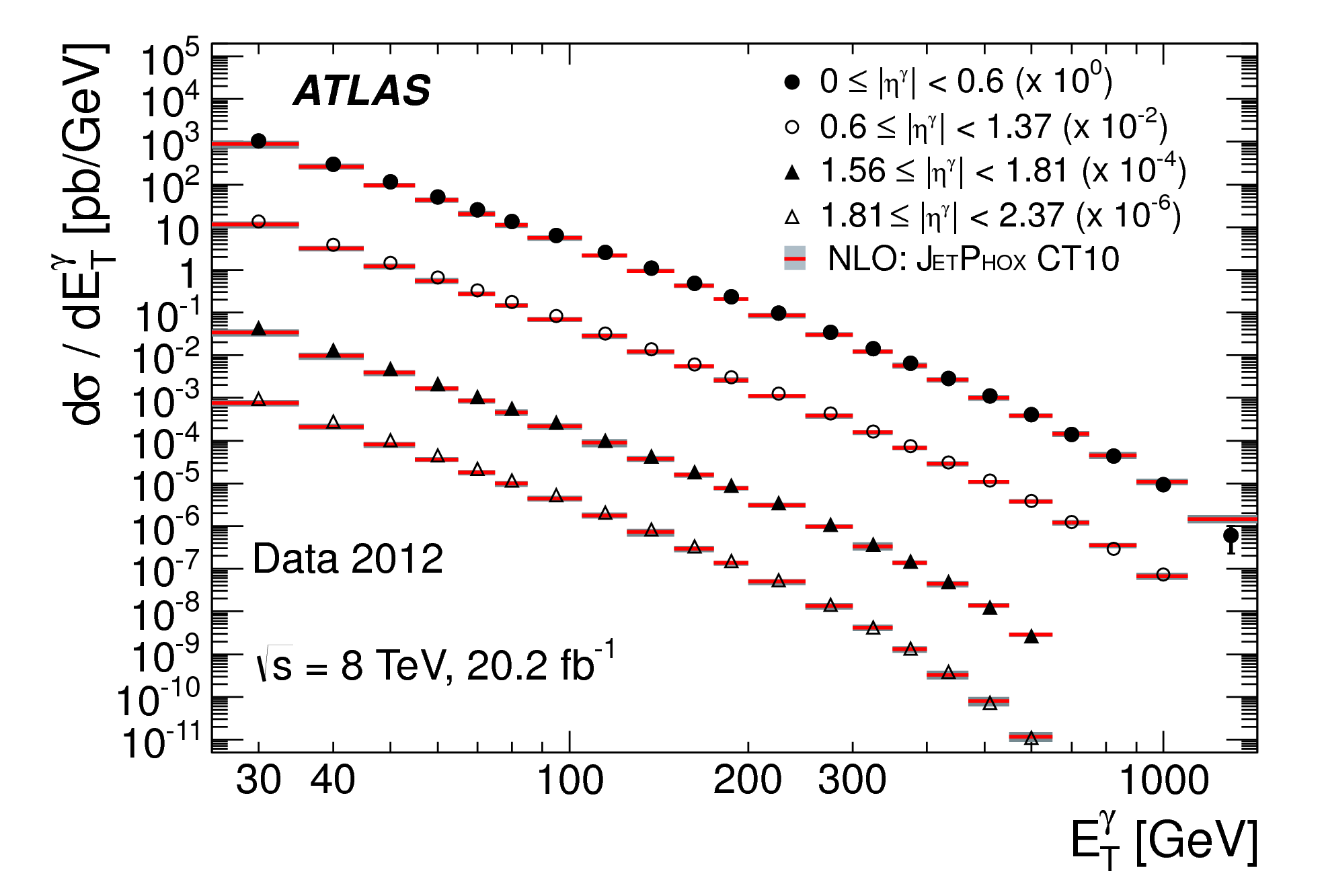}
\caption[Photon cross-section at $\sqrt{s}$=8 TeV - comparison of data and Monte Carlo]{Cross-section of prompt photon production at \mbox{$\sqrt{s}$=8 TeV} \cite{Aad:2016xcr} - comparison of data and Monte Carlo (\textsc{JetPhox}). $\eta^{\gamma}$ range is split into 4 bins ($0\leq |\eta^{\gamma}|<0.6$, $0.6\leq |\eta^{\gamma}|<1.37$, $1.56\leq |\eta^{\gamma}|<1.81$, and $1.81\leq |\eta^{\gamma}|<2.37$). Both statistical and systematic uncertainties are included.}\label{fig_ph_8tev_cs}
\end{figure}
\begin{figure}
\centering
\begin{minipage}{.88\textwidth}
  \centering
  \includegraphics[width=\linewidth]{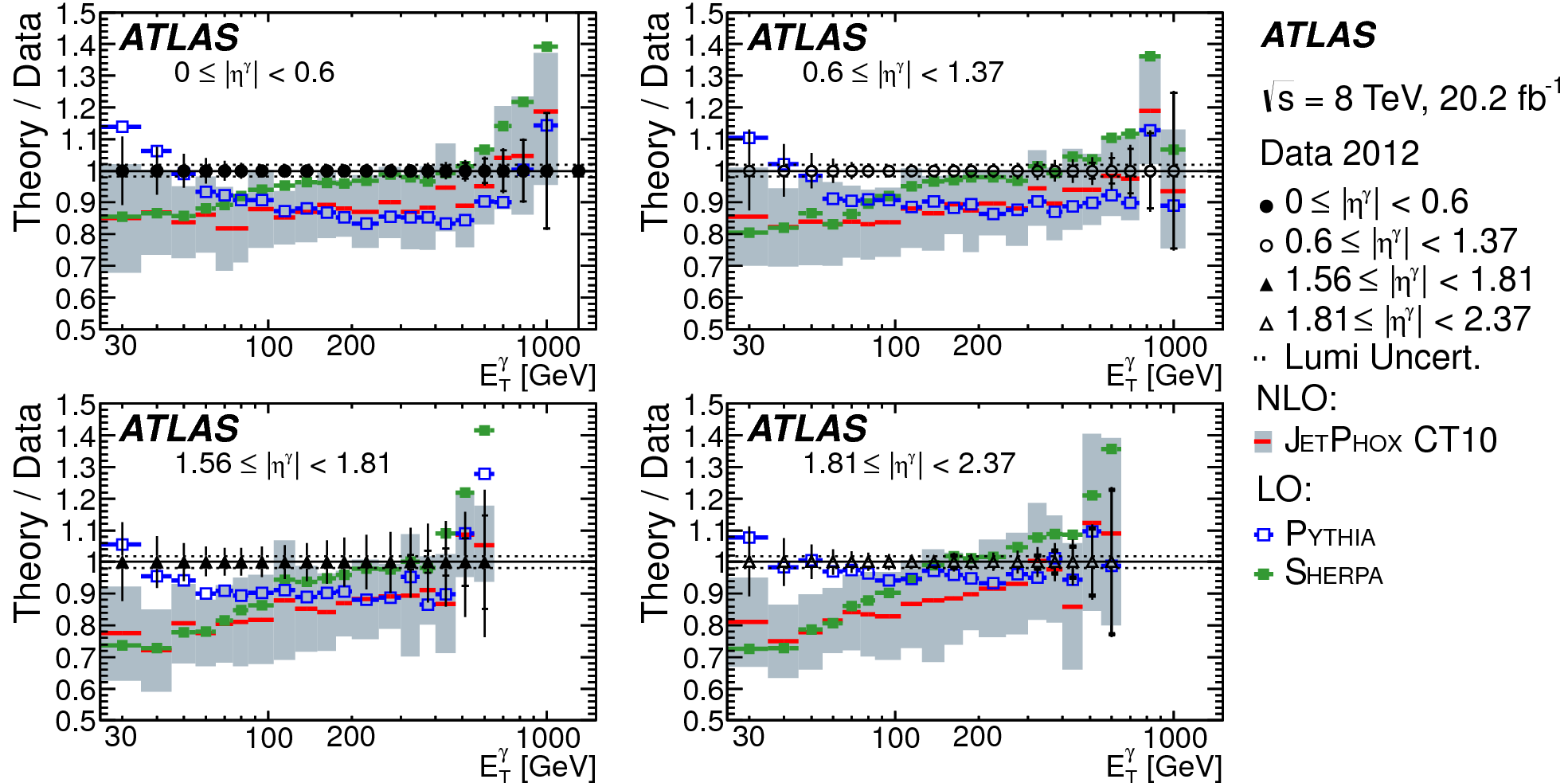}
\end{minipage}%
\\
\begin{minipage}{.88\textwidth}
  \centering
  \includegraphics[width=\linewidth]{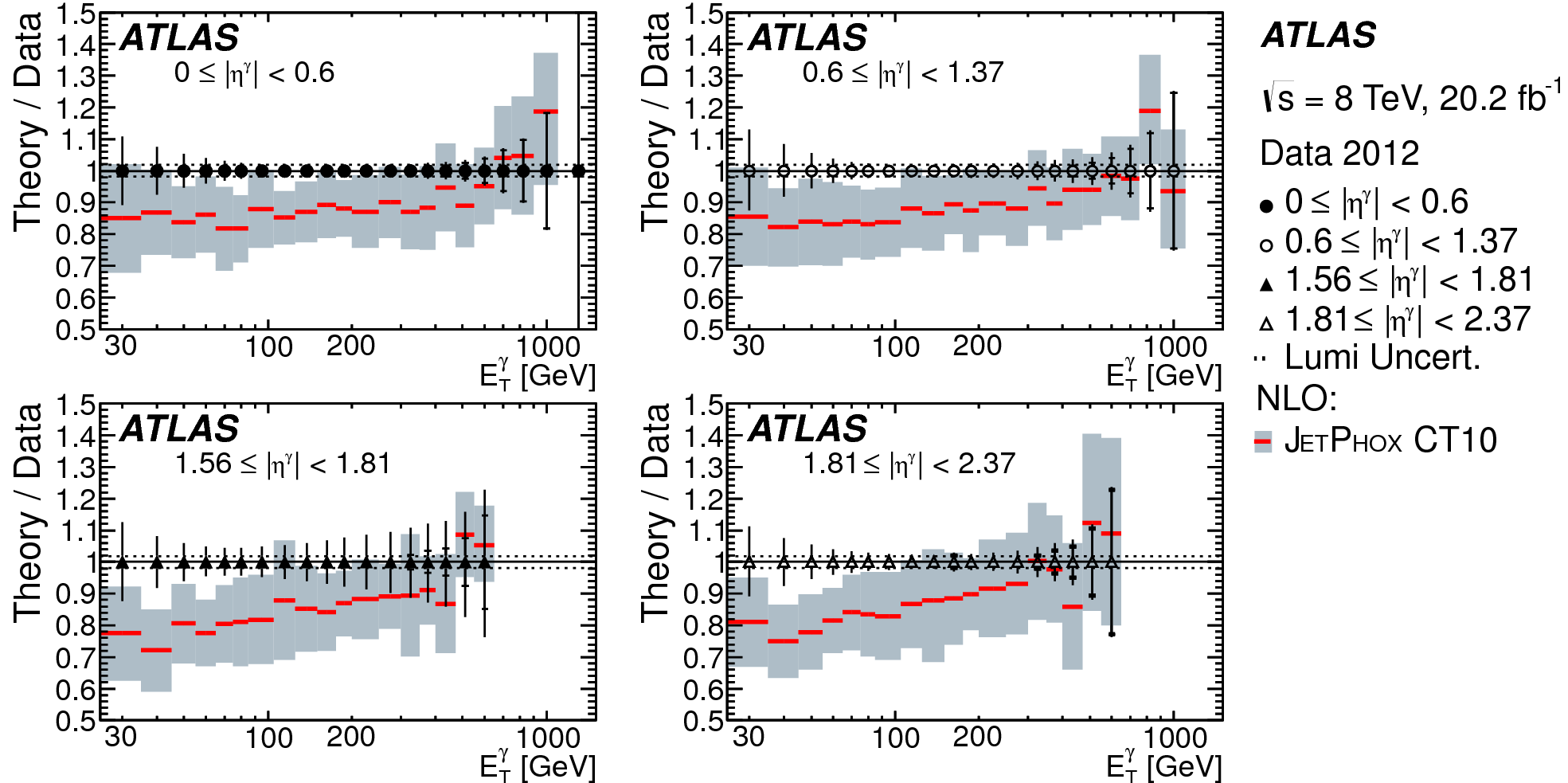}
\end{minipage}%
\\
\begin{minipage}{.88\textwidth}
  \centering
  \includegraphics[width=\linewidth]{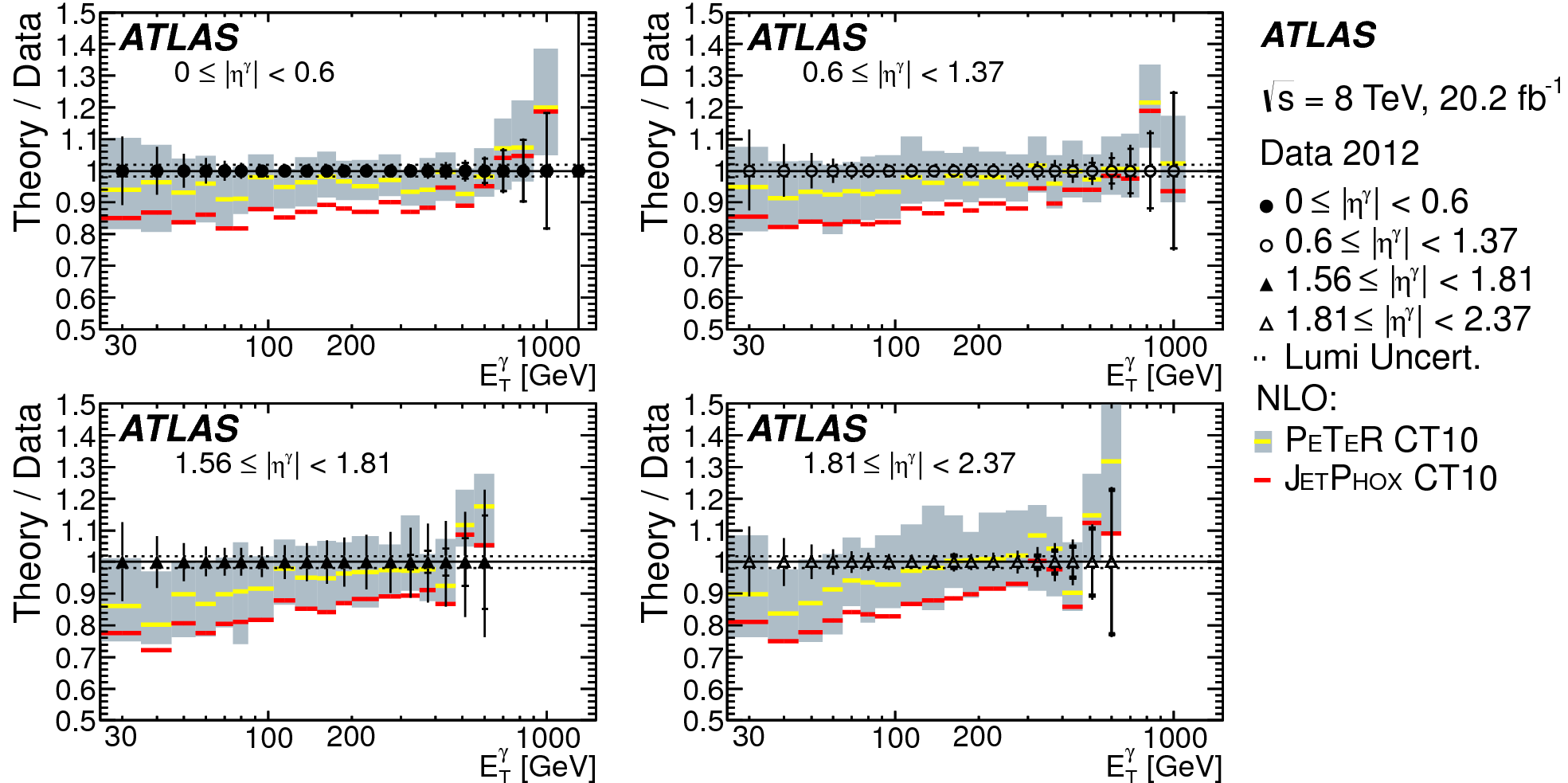}
\end{minipage}%
\caption[Photon at $\sqrt{s}$=8 TeV - comparison of data and Monte Carlo]{Ratios Theory/Data of prompt photon cross-section at \mbox{$\sqrt{s}$=8 TeV} \cite{Aad:2016xcr}. The top four plots correspond to the LO Monte Carlo (\textsc{Pythia} 8.165 and \textsc{Sherpa} 1.4.0) and to the NLO Monte Carlo \textsc{JetPhox}. Middle four plots correspond to the NLO \textsc{JetPhox} Monte Carlo only. Ratios obtained from both NLO generators \textsc{PeTer} and \textsc{JetPhox} are shown in the bottom four plots - uncertainies correspond to \textsc{PeTer} only. $\eta^{\gamma}$ range is split into 4 bins ($0\leq |\eta^{\gamma}|<0.6$, $0.6\leq |\eta^{\gamma}|<1.37$, $1.56\leq |\eta^{\gamma}|<1.81$, and $1.81\leq |\eta^{\gamma}|<2.37$). Both statistical and systematic uncertainties are included.}\label{fig_ph_8tev_comp}
\end{figure}
\newline\indent The cross-section of prompt photon production is shown in Figure \ref{fig_ph_8tev_cs} for data and the NLO Monte Carlo \textsc{JetPhox}. The Monte Carlo describes the shape of the data well over ten orders of magnitude. The agreement is similar for all tested PDF sets (CT10, MSTW2008, NNPDF2.3, and HeraPDF1.5).
\newline\indent Ratios theory/data for LO Monte Carlo (\textsc{Pythia} and \textsc{Sherpa}) are shown in the four top plots of Figure \ref{fig_ph_8tev_comp}. \textsc{Sherpa} matches the data in $100\leq E_{T}^{\gamma}<500$ GeV. In the lower $E_{T}^{\gamma}$ region (where a larger fragmentation contribution is expected), \textsc{Sherpa} follows the predictions from \textsc{JetPhox}. In the high-$E_{T}^{\gamma}$ region, it tends to go above the measured value. \textsc{Pythia} overestimates the measured cross-section in the low-$E_{T}^{\gamma}$ region (the fragmentation contribution is not well modeled by the parton shower). In the rest, it is similar to \textsc{JetPhox}. The ratios are shown in the bottom eight plots of Figure \ref{fig_ph_8tev_comp} for the NLO Monte Carlo (\textsc{PeTer} and \textsc{JetPhox}). Predictions of \textsc{JetPhox} are lower than data (up to 20\%) but still within uncertainties. The description of the data by \textsc{PeTer} Monte Carlo is much better.

\section{Photon measurement at $\sqrt{s}$=13 TeV}
The isolated prompt photon measurement at \mbox{$\sqrt{s}$=13 TeV} \cite{ATL-PHYS-PUB-2015-016} used \mbox{6.4 pb${}^{-1}$} of data collected in 2015.
\begin{figure}[h]
\centering
\begin{minipage}{.5\textwidth}
  \centering
  \includegraphics[width=\linewidth]{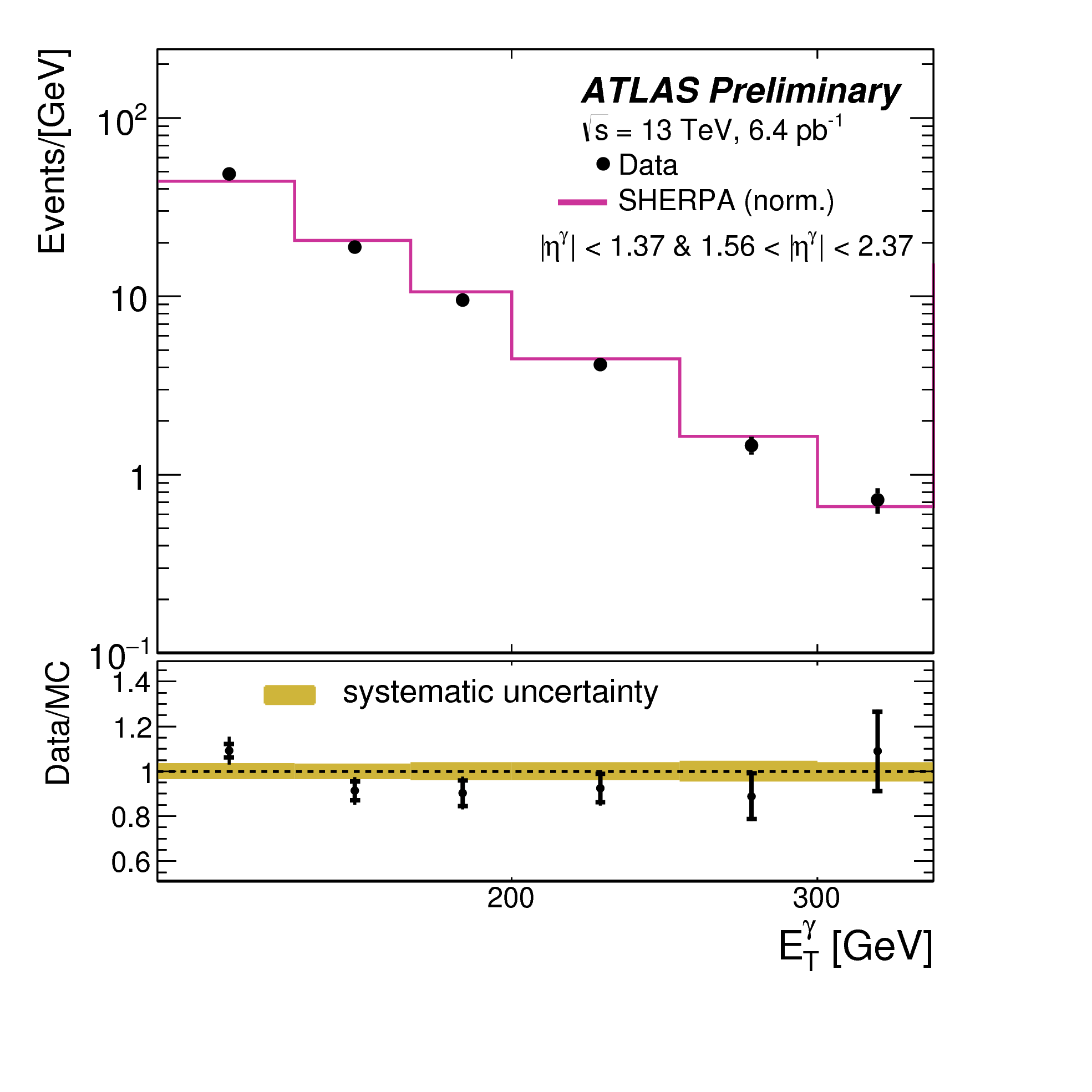}
\end{minipage}%
\begin{minipage}{.45\textwidth}
  \centering
  \includegraphics[width=\linewidth]{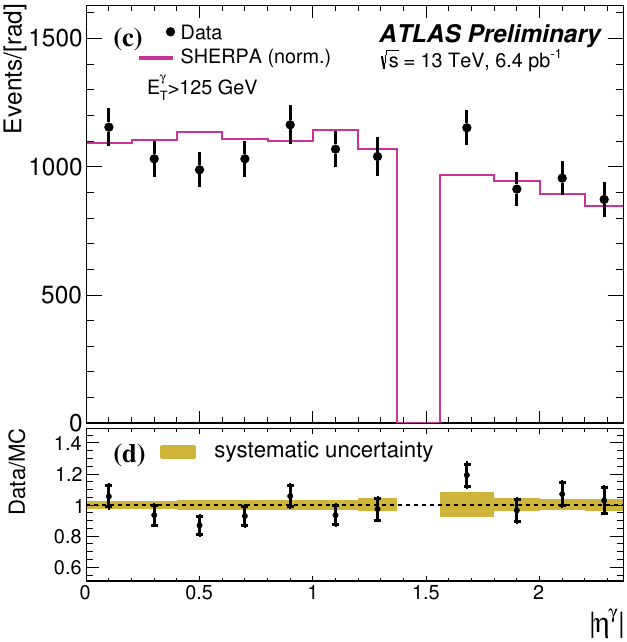}
\end{minipage}%
\caption[Photon at $\sqrt{s}$=13 TeV - comparison of yield in data and LO Monte Carlo]{Distributions of event
yield as a function of $E_{T}^{\gamma}$ and $\eta^{\gamma}$ for data and \textsc{Sherpa} at $\sqrt{s}$=13 TeV \cite{ATL-PHYS-PUB-2015-016}. Error bars represent statistical uncertainty and yellow band systematic uncertainty.}\label{fig_ph_13tev}
\end{figure}
\newline\indent The selected events are required to have at least one reconstructed primary vertex, with at least two associated tracks, consistent with the average beam-spot position. Photons are required to have $|\eta^{\gamma}|<1.37$, $1.56\leq |\eta^{\gamma}|<2.37$ and $E_{T}^{\gamma}>125$ GeV. They are triggered using a single photon trigger with threshold 120 GeV. Tight isolated photons are considered as signal also in this analysis.
\newline\indent The background is subtracted using a 2D sideband method (described in section \ref{sec_ph_8tev}). Additional background (e.g. from multijet processes) removal is required. The events are divided into two groups, one with tight photons and one with non-tight photons. The non-tight $E_{T}^{iso}$ distribution is normalised such that the integrals of the tight and non-tight distributions in the range $10 < E_{T}^{iso} < 25$ GeV coincide. Contribution of non-tight photons in isolated photons region is considered to be a background and is subtracted to obtain the final signal yield.
\newline\indent Data are compared to LO Monte Carlo. The simulations are done by Sherpa 2.1.1 using CT10 PDF.
\newline\indent Systematic uncertainties are dominated by the photon energy scale and resolution, by the identification and trigger efficiency, and by the modeling of isolation energy. A systematic uncertainty of 5\% is assigned to the 2D sideband method.
\newline\indent Distributions of the event yield as a function of $E_{T}^{\gamma}$ and $\eta^{\gamma}$ for data and \textsc{Sherpa} are shown in Figure \ref{fig_ph_13tev}. The description of the data by the Monte Carlo is good.

\section{Jet reconstruction and identification}
Jets are reconstructed from topological clusters of cells in the calorimeter using the anti-kT algorithm with radius 0.4. They are calibrated using transverse momentum and pseudorapidity dependent corrections from Monte Carlo. Jets in data are additionally corrected by a factor based on in situ studies.
\newline\indent Identification criteria are selected to reject fake jets reconstructed from non-collision signals - like beam-related background, cosmic rays, or detector noise, etc.

\section{Jet cross-section measurement at $\sqrt{s}$=13 TeV}
The inclusive jet cross-section measurement \cite{ATLAS-CONF-2015-034} used \mbox{78 pb${}^{-1}$} of data collected in 2015 at \mbox{$\sqrt{s}$=13 TeV}.
\newline\indent The selected events are required to have at least one well-reconstructed vertex, which must have at least two associated tracks with transverse momentum greater than 400 MeV and must be consistent with the beam spot of the proton-proton collisions. Jets are measured in the rapidity range $\eta<0.5$ and with $346<p_{T}<838$ GeV. The rapidity range is chosen to be within the coverage of the barrel calorimeters. The upper $p_{T}$ limit is caused by the fact that the performance of the detector for jets with higher $p_{T}$ is still under study. Single jet triggers with various thresholds are used. Jets are also required to pass a looser identification criterion. 
\newline\indent The data are compared to NLO Monte Carlo. The simulations are done by NLOJET++ 4.1.3 \cite{Nagy:2003tz} (interfaced to APPLGRID \cite{Carli:2010rw}) using CT10, MMHT and NNPDF 3.0 PDFs.
\begin{figure}
\centering
\begin{minipage}{.35\textwidth}
  \centering
  \includegraphics[width=\linewidth]{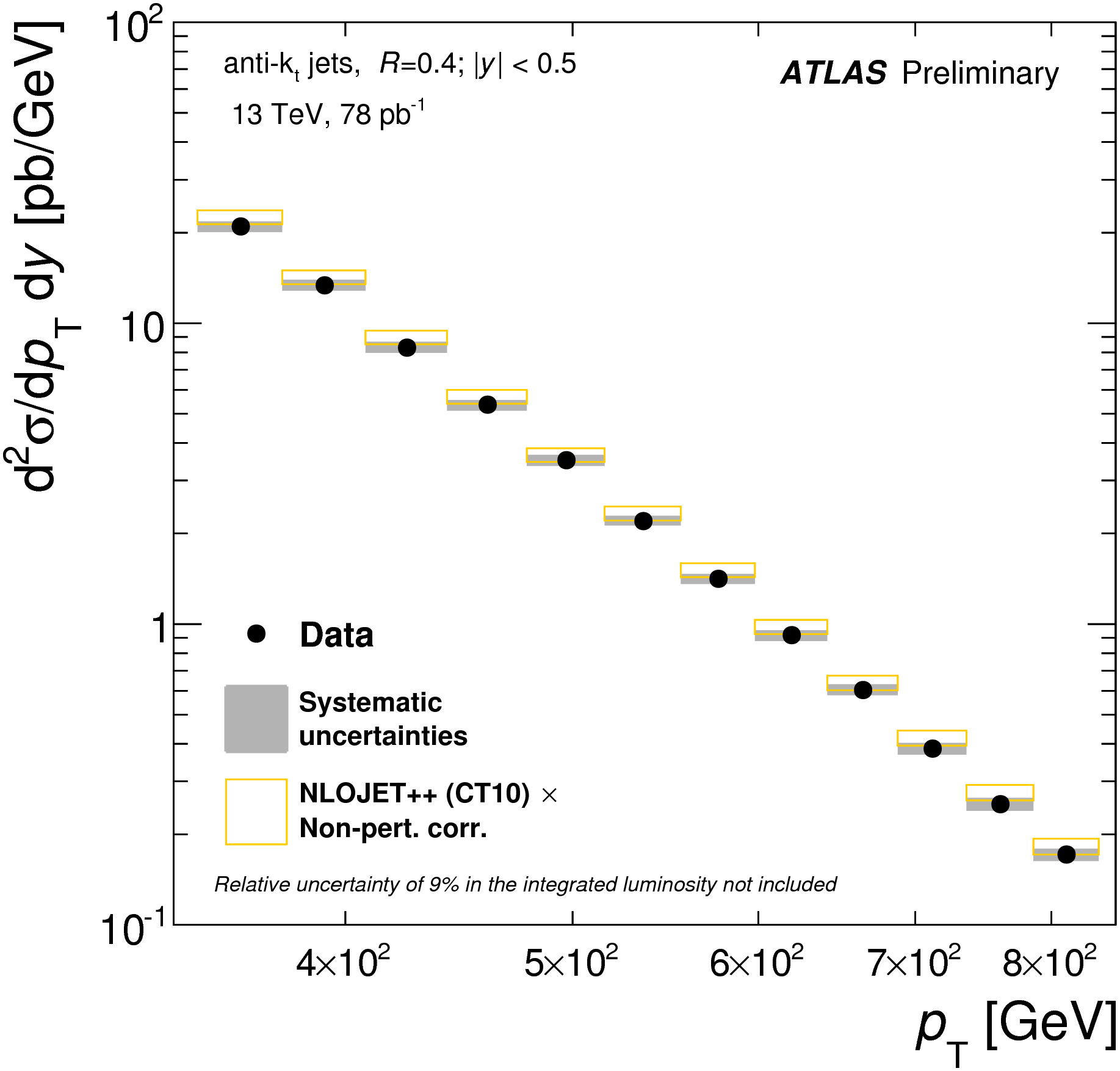}
\end{minipage}%
\begin{minipage}{.6\textwidth}
  \centering
  \includegraphics[width=\linewidth]{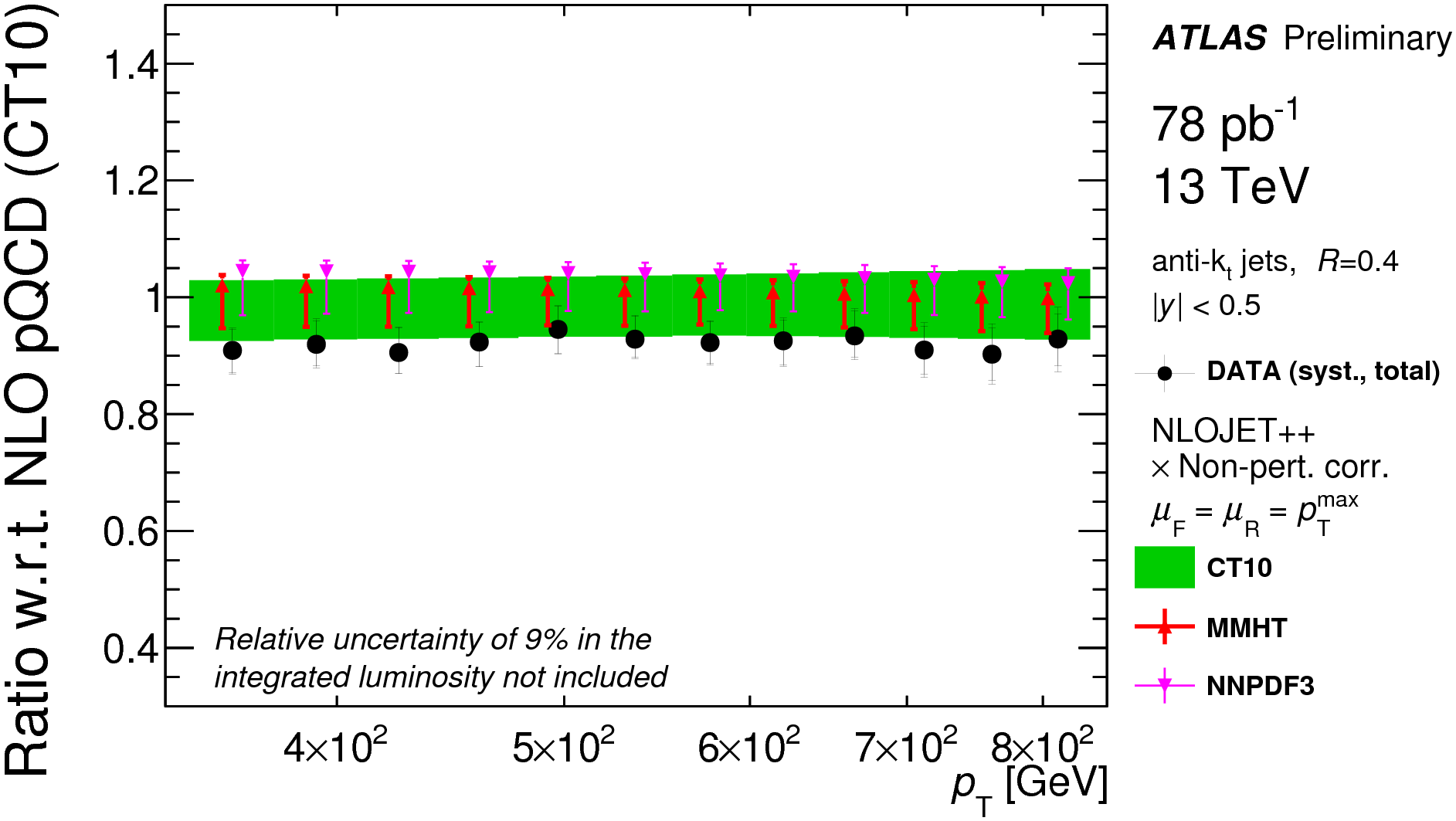}
\end{minipage}%
\caption[Comparison of data and NLO Monte Carlo in inclusive jet at $\sqrt{s}$=13 TeV]{Comparison of data and NLO Monte Carlo of jet measurement at \mbox{$\sqrt{s}$=13 TeV} \cite{ATLAS-CONF-2015-034}. Left plot shows cross-section with error bars representing statistical uncertainties and grey band representing systematic uncertainties. Right plot shows comparison of data to NLO using several different PDFs. Inner error bars represent systematic uncertainties, outer bars total uncertainties, and green band uncertainties in the prediction.}\label{fig_j_13tev}
\end{figure}
\newline\indent The systematic uncertainty is dominated by the jet energy scale and by the resolution and unfolding uncertainties. The luminosity uncertainty is 9\%.
\newline\indent The results are in Figure \ref{fig_j_13tev}.  NLO prediction (using CT10 PDF) describes data well over two orders of magnitude. The dependence of Monte Carlo prediction on PDF used (MMHT and NNPDF3 were tested) is weak.

\section{Summary}
The cross-section of isolated prompt photon is measured using \mbox{20.2 fb${}^{-1}$} of data collected at \mbox{$\sqrt{s}$=8 TeV} in 2012. It is measured as a function of transverse energy of the leading photon in four pseudorapidity bins ($0\leq |\eta^{\gamma}|<0.6$, $0.6\leq |\eta^{\gamma}|<1.37$, $1.56\leq |\eta^{\gamma}|<1.81$, and $1.81\leq |\eta^{\gamma}|<2.37$). Data are compared to LO Monte Carlo (\textsc{Pythia} and \textsc{Sherpa}) and NLO Monte Carlo (\textsc{PeTer} and \textsc{JetPhox}). \textsc{JetPhox} describes the shape of the data well over ten orders of magnitude. Its predictions are lower than data (up to 20\%) but still within uncertainties. This is valid for all tested PDFs. The description of the data by the \textsc{PeTer} Monte Carlo is much better because of its beyond NLO calculations. \textsc{Sherpa} shows a different behaviour in three different regions when compared to data. In the low-$E_{T}$ region, \textsc{Sherpa} follows the predictions from \textsc{JetPhox}. A larger fragmentation contribution is expected there. In the region $100\leq E_{T}^{\gamma}<500$ GeV, \textsc{Sherpa} matches the data. In the high-$E_{T}^{\gamma}$ region, it rises above the measured value. \textsc{Pythia} overestimates data in the low-$E_{T}^{\gamma}$ region. The fragmentation contribution is not well modeled by the parton shower in this area. In the rest of the $E_{T}^{\gamma}$ range, \textsc{Pythia} is similar to \textsc{JetPhox}.
\newline\indent The photon yield is measured at \mbox{$\sqrt{s}$=13 TeV} using \mbox{78 pb${}^{-1}$} of data collected in 2015. It is measured as a function of transverse energy and pseudorapidity of the leading photon. The data are compared to \textsc{Sherpa}. The description of the data by the Monte Carlo is good.
\newline\indent The differential cross-section of the leading jet as a function of jet transverse momentum is measured using \mbox{78 pb${}^{-1}$} of data collected at \mbox{$\sqrt{s}$ = 13 TeV} in 2015. NLOJET++ 4.1.3 using CT10 prediction describes data well over two orders of magnitude.

\end{document}